# Joint Probabilities of Photon polarization Correlations in $e^+e^-$ Annihilation

N. Yongram and E. B. Manoukian

Joint Probability distributions of photon polarization *correlations* are computed, as well as those corresponding to the cases when only one of the photon polarizations are measured, in $e^+e^-$ annihilation in *flight* in QED. This provides a *dynamical*, rather than a kinematical, description of photon polarization correlations as stemming from the ever precise and realistic QED theory. Such computations may be relevant to recent and future experiments involved in testing Bell-like inequalities as described.

1. INTRODUCTION

The purpose of this paper is to derive the explicit joint probability distributions of photon ($\gamma\gamma$) polarization *correlations* in $e^+e^-$ annihilation in *flight* [1] in QED, as well as to obtain the corresponding probabilities when only one of the photon's polarization is measured, This provides clear cut *dynamical*, rather than kinematical, descriptions of photon polarizations correlations as follow directly from this monumental and experimentally reliable QED theory. Particle correlations have been systematically studied earlier [e.g.,1-4] emphasizing, however, different experimental situations and aspects. Polarizations phenomenae were studied many years ago [5], we are, however, interested in correlations aspects which have been quite important experimentally in recent years [6-9] in the light of the foundations of quantum physics vis-à-vis Bell-like inequalities. Two types of collisions are considered for $e^+e^-$ annihilation in flight in the c.m. motion. The first one in which a $e^+$ and an $e^-$ in the c.m., initially prepared to be moving along a specific axis, annihilate each other and two photons are observed to be moving along a given specific axis. Given that this process has occurred, we compute the conditional joint probability distributions of photon polarizations as well as the probabilities corresponding to the measurement of only one of the photon's polarization. The second one is involved with all repeated experiments corresponding to all orientations of the axis of motion of $e^+e^-$ pairs in the c.m. initially prepared with the same speeds, and a pair of photons is observed moving along a given axis in each case after the annihilation process. Given that these collisions have accurred, we compute the conditional probabilities of photon polarizations correlations mentioned above. In this latter case we must average over the initial orientations of the axis along which a $e^+e^-$ pair may initially move before annihilation occurs. With the explicit expressions for the probabilities derived from this quantum *dynamical* analysis, we finally show a clear violation of the relevant Bell-like inequality [6-9] as a function of the speed of $e^+$ (or of $e^-$). Our convention for the metric is $[g_{\mu\nu}] = diag[-1,1,1,1]$.

## 2. COMPUTATIONS OF THE PROBABILITY DISTRIBUTIONS

The transition probability of $e^+(p_1) e^-(p_2) \to \gamma(k_1) \gamma(k_2)$ to the leading order in the fine-structure constant $\alpha$ is, up to an unimportant multiplicative factor for the problem at hand, given by [e.g., 10, 11]

$$P_r \propto \left[ \frac{1}{4} \frac{(k_1 k_2)^2}{(p_1 k_1)(p_1 k_2)} - \left( \varepsilon_{1\lambda} \cdot \varepsilon_{2\lambda'} \right)^2 \right] \tag{1}$$

where

$$\varepsilon_{1\lambda}^\mu = \left( \delta^\mu{}_\nu - \frac{p_{1\nu} k_1^\mu}{p_1 k_1} \right) e^\mu{}_1(\lambda), \quad k_1 e_1(\lambda) = 0 \tag{2}$$

$$\varepsilon_{1\lambda'}^\mu = \left( \delta^\mu{}_\nu - \frac{p_{1\nu} k_2^\mu}{p_1 k_2} \right) e^\mu{}_2(\lambda'), \quad k_2 e_2(\lambda') = 0 \tag{3}$$

$e_{1,2}^\nu(\lambda)$ Denote the polarization vectors of the photons satisfying the completeness relation

$$\sum_\lambda e^\mu{}_i(\lambda) e^\nu{}_i(\lambda) = g^{\mu\nu} - \frac{k_i^\mu \bar{k}_i^\nu + \bar{k}_i^\mu k_i^\nu}{k_i \bar{k}_i} \tag{4}$$

(no sum over $i$), where $k = (k^0, \vec{k})$, $\bar{k} = (k^0, -\vec{k})$. We note that $\varepsilon_{i\lambda}^\mu$ are invariant under the gauge transformations $e_i^\nu(\lambda) \to e_i^\nu(\lambda) + k_i^\nu b_\lambda(k_i)$ for arbitrary $b_\lambda(k_i)$.

In the c.m. of a pair $e^+e^-$

$$\left.\begin{array}{c} \vec{p}_2 = -\vec{p}_1 \equiv -\vec{p}, \quad \vec{k}_2 = -\vec{k}_1 \equiv -\vec{k}, \quad p_1^0 = p_2^0 = k_1^0 = k_2^0 \equiv p^0 \\ k^0 = |\vec{k}|, \quad p^0 = \sqrt{\vec{p}^2 + m^2} \end{array}\right\} \quad (5)$$

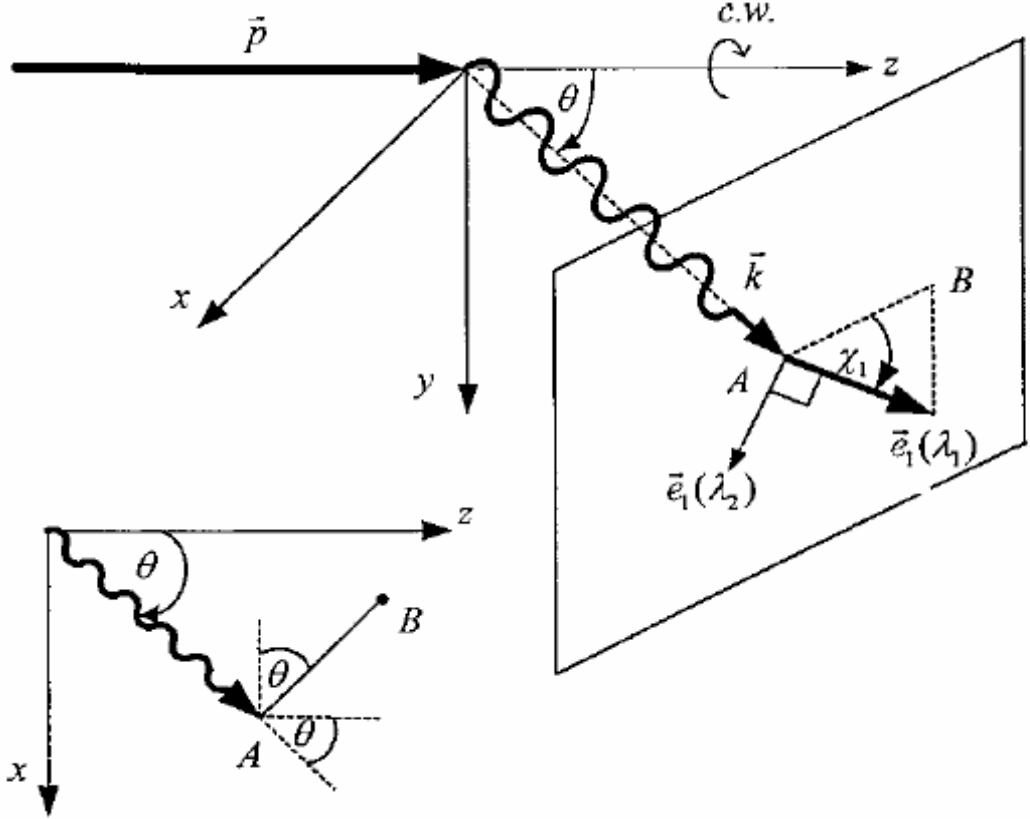

**Fig. 1.** In this figure $\vec{k}$ lies in the $x-z$ plane and $\vec{p}$ is along the $z$-axis. The polarizations vectors $\vec{e}_1(\lambda_1)$, $\vec{e}_1(\lambda_2)$ are orthogonal to each other and are orthogonal to $\vec{k}$. The line segment $AB$, of length $|\cos \chi_1|$, lies in the $x-z$ plane. By rotating the coordinate system c.w. about the $z$-axis, by an angle $\phi$, the vectors $\vec{k}$, $\vec{e}_1(\lambda_1)$, $\vec{e}_1(\lambda_2)$ will have general orientations in the resulting coordinate system.

In the **Fig. 1.** we show how to introduce the polarization $e_1^\mu(\lambda) = (0, \vec{e}_1(\lambda))$ in reference to the vector $k^\mu$. If $\vec{k}$ is chosen to lie in the $x-z$ plane, then

$$\vec{k} \; : \; |\vec{k}|(\sin\theta, 0, \cos\theta) \tag{6}$$

and from the figure, with $\vec{e}_1(\lambda) \equiv \vec{e}_1$,

$$\vec{e}_1 \; : \; (-\cos\theta\cos\chi_1, \sin\chi_1, \sin\theta\cos\chi_1) \tag{7}$$

where, here, $\vec{p} = |\vec{p}|(0,0,1)$. For a general orientation of $\vec{k}$ and $\vec{e}_1$, we must rotate the $x-y-z$ coordinate system c.w. about the $z$–axis by an angle $\phi$. This is accomplished by the rotation matrix $R$ with matrix elements:

$$R^{il} = \delta^{il} + \varepsilon^{ijl}\frac{p^j}{|\vec{p}|}\sin\phi + \left(\delta^{il} - \frac{p^i p^l}{|\vec{p}|^2}\right)(\cos\phi - 1) \tag{8}$$

giving

$$\vec{k} = |\vec{k}|(\cos\phi\sin\theta, \sin\phi\sin\theta, \cos\theta) \tag{9}$$

as expected, and

$$\vec{e}_1 = (-\cos\theta\cos\chi_1\cos\phi - \sin\chi_1\sin\phi, \sin\chi_1\cos\phi - \cos\theta\cos\chi_1\sin\phi, \sin\theta\cos\chi_1) \tag{10}$$

in the resulting coordinate system. A similar expression for $\vec{e}_2(\lambda') \equiv \vec{e}_2$ is obtained by replacing $\chi_1$ by $\chi_2$. With $\vec{e}_1 \equiv \vec{e}_1(\lambda_1)$, $\vec{e}_2(\lambda_2)$ is obtained from $\vec{e}_1$ by the substitution $\chi_1 \to \chi_1 + \pi/2$.

In the c.m. of $e^+e^-$, (1) may rewritten in the convenient form

$$P_r \propto \left[ \frac{1}{4} \frac{(k_1 k_2)^2}{(p_1 k_1)(p_1 k_2)} - \left( \vec{e}_1 \cdot \vec{e}_2 + \frac{\vec{e}_1 \cdot \vec{p}\, \vec{e}_2 \cdot \vec{p}(k_1 k_2)}{(p_1 k_1)(p_1 k_2)} \right)^2 \right] \tag{11}$$

We treat two processes of annihilation associated with the relative probability given in (11).

*Process 1:*

We consider the annihilation of $e^+ e^-$ pairs in the flight in the c.m. (located at the origin of the coordinate system) initially prepared to be moving along the $z$–axis, as in the figure, each moving with speed $v = \beta c$, prior to their annihilation into pairs of photons, and place detectors for the latter at opposite ends of the $x$–axis.

Using the scalar products

$$\vec{e}_i \cdot \vec{p} = |\vec{p}| \sin\theta \cos\chi_i, \quad \vec{p} \cdot \vec{k}_1 = |\vec{p}||\vec{k}| \cos\theta = -\vec{p} \cdot \vec{k}_2 \tag{12}$$

we obtain by a direct evaluation of (11)

$$P_r \propto \frac{[1 - 4(1-\beta^2)\cos\chi_1 \cos\chi_2 (\cos(\chi_1 - \chi_2) - 2\cos\chi_1 \cos\chi_2)]}{(1 - \beta^2 \cos^2\theta)}$$

$$- \frac{4(1-\beta^2)^2 \cos^2\chi_1 \cos^2\chi_2}{(1-\beta^2 \cos^2\theta)^2} - [\cos(\chi_1 - \chi_2) - 2\cos\chi_1 \cos\chi_2]^2$$

(13)

where $\beta = |\vec{p}|/p^0$ is the speed of $e^+$ (or of $e^-$) divided by the speed of light, and $\theta$ is the angle between $\vec{k}$ and $\vec{p}$. We note that the angles $\chi_1$, $\chi_2$ have given fixed values when the vector $\vec{k}$ is made to rotate in the coordinate system.

Since $\theta$ is a continuous variable, we may integrate the expression in (13) over $\theta$ from $\pi/2-\delta$ to $\pi/2+\delta$ and then rigorously take the limit $\delta \to 0$ in evaluating the normalized probabilities in question. The $\phi$–integral, here, is not important in evaluating these normalized probabilities since it leads to overall multiplicative factors which cancel out in the final expressions.

Upon using the integrals

$$\int_{\pi/2-\delta}^{\pi/2+\delta} \frac{\sin\theta d\theta}{(1-\beta^2 \cos^2\theta)} = \frac{1}{\beta}\ln\left(\frac{1+\beta\sin\delta}{1-\beta\sin\delta}\right) \tag{14}$$

$$\int_{\pi/2-\delta}^{\pi/2+\delta} \frac{\sin\theta d\theta}{(1-\beta^2 \cos^2\theta)^2} = \frac{1}{\beta}\left[\frac{\beta\sin\delta}{1-\beta^2 \sin^2\delta} + \frac{1}{2}\ln\left(\frac{1+\beta\sin\delta}{1-\beta\sin\delta}\right)\right] \tag{15}$$

we obtain from (13)

$$\int_{\pi/2-\delta}^{\pi/2+\delta} \sin\theta d\theta P_r \propto \left\{ \frac{[1-4(1-\beta^2)\cos\chi_1 \cos\chi_2(\cos(\chi_1-\chi_2)-2\cos\chi_1 \cos\chi_2)]}{\beta}\ln\left(\frac{1+\beta\sin\delta}{1-\beta\sin\delta}\right) \right.$$

$$-4(1-\beta^2)^2 \cos^2\chi_1 \cos^2\chi_2 \left[\frac{\sin\delta}{1-\beta^2 \sin^2\delta} + \frac{1}{2\beta}\ln\left(\frac{1+\beta\sin\delta}{1-\beta\sin\delta}\right)\right]$$

$$\left. -2\sin\delta[\cos(\chi_1-\chi_2)-2\cos\chi_1 \cos\chi_2]^2 \right\} \equiv F_\delta(\chi_1,\chi_2)$$

(16)

To normalize the expression in (16), we have sum $F_\delta(\chi_1,\chi_2)$ over the polarizations directions specified by the pairs of angles:

$$(\chi_1,\chi_2),(\chi_1+\pi/2,\chi_2),(\chi_1,\chi_2+\pi/2),(\chi_1+\pi/2,\chi_2+\pi/2) \tag{17}$$

That is, we have to find the normalization factor

$$N_\delta = F_\delta(\chi_1, \chi_2) + F_\delta(\chi_1 + \pi/2, \chi_2)$$
$$+ F_\delta(\chi_1, \chi_2 + \pi/2) + F_\delta(\chi_1 + \pi/2, \chi_2 + \pi/2) \tag{18}$$

The latter works out to be

$$N_\delta = [4 - 4(1-\beta^2) - 2(1-\beta^2)^2]\frac{1}{\beta}\ln\left(\frac{1+\beta\sin\delta}{1-\beta\sin\delta}\right) - 4(1-\beta^2)^2\frac{\sin\delta}{1-\beta^2\sin^2\delta}$$
$$- 4\sin\delta \tag{19}$$

Therefore, given that the process has occurred as described above, with two photons moving (back-to-back) along the $x$–axis, the conditional joint probability of the polarizations, specified by the angles $\chi_1, \chi_2$ is rigorously given by

$$P(\chi_1, \chi_2) = \lim_{\delta \to 0} \frac{F_\delta(\chi_1, \chi_2)}{N_\delta} \tag{20}$$

For all $0 \le \beta \le 1$, we use the limit

$$\frac{1}{\beta}\ln\left(\frac{1+\beta\sin\delta}{1-\beta\sin\delta}\right) \underset{\delta \to 0}{\widetilde{\phantom{xx}}} 2\delta \tag{21}$$

to obtain from (16), (19), (20),

$$P(\chi_1, \chi_2) = \frac{1 - (\cos(\chi_1 - \chi_2) - 2\beta^2 \cos\chi_1 \cos\chi_2)^2}{2[1 + 2\beta^2(1-\beta^2)]} \tag{22}$$

for all $0 \le \beta \le 1$.

If only one of the polarizations is measured, the we have to evaluate $F_\delta(\chi_1,\chi_2) + F_\delta(\chi_1,\chi_2+\pi/2)$ and $F_\delta(\chi_1,\chi_2) + F_\delta(\chi_1+\pi/2,\chi_2)$. To this end, (16) gives

$$F_\delta(\chi_1,\chi_2) + F_\delta(\chi_1,\chi_2+\pi/2) = [2 + 2(1-\beta^4)\cos^2\chi_1]\frac{1}{\beta}\ln\left(\frac{1+\beta\sin\delta}{1-\beta\sin\delta}\right)$$

$$-4(1-\beta^2)^2\frac{\sin\delta}{1-\beta^2\sin^2\delta} - 2\sin\delta$$

(23)

$$F_\delta(\chi_1,\chi_2) + F_\delta(\chi_1+\pi/2,\chi_2) = [2 + 2(1-\beta^4)\cos^2\chi_2]\frac{1}{\beta}\ln\left(\frac{1+\beta\sin\delta}{1-\beta\sin\delta}\right)$$

$$-4(1-\beta^2)^2\frac{\sin\delta}{1-\beta^2\sin^2\delta} - 2\sin\delta$$

(24)

That is, the conditional probabilities associated with the measurement of only of the polarizations are given by

$$P(\chi_1,\_) = \lim_{\delta\to 0}\frac{F_\delta(\chi_1,\chi_2) + F_\delta(\chi_1,\chi_2+\pi/2)}{N_\delta} \tag{25}$$

$$P(\_,\chi_2) = \lim_{\delta\to 0}\frac{F_\delta(\chi_1,\chi_2) + F_\delta(\chi_1+\pi/2,\chi_2)}{N_\delta} \tag{26}$$

Form (23)-(26), and (18), these work out to be simply given by

$$P(\chi_1,\_) = \frac{1 + 4\beta^2(1-\beta^2)\cos^2\chi_1}{2[1+2\beta^2(1-\beta^2)]} \tag{27}$$

$$P(\_,\chi_2) = \frac{1+4\beta^2(1-\beta^2)\cos^2\chi_2}{2[1+2\beta^2(1-\beta^2)]}$$

(28)

for all $0 \leq \beta \leq 1$, and are, respectively, *dependent* on $\chi_1, \chi_2$.

We note the important statistical property that

$$P(\chi_1,\chi_2) \neq P(\chi_1,\_)P(\_,\chi_2) \qquad (29)$$

in general.

In the notation of Local Hidden Variables (LHV) theory [6-9], we have the identifications:

$$P(\chi_1,\chi_2) = \frac{P_{12}(a_1,a_2)}{P_{12}(\infty,\infty)} \qquad (30)$$

$$P(\chi_1,\_) = \frac{P_{12}(a_1,\infty)}{P_{12}(\infty,\infty)} \qquad (31)$$

$$P(\_,\chi_2) = \frac{P_{12}(\infty,a_2)}{P_{12}(\infty,\infty)} \qquad (32)$$

Defining

$$S = P(\chi_1,\chi_2) - P(\chi_1,\chi_2') + P(\chi_1',\chi_2) + P(\chi_1',\chi_2') - P(\chi_1',\_) - P(\_,\chi_2) \qquad (33)$$

for four angles $\chi_1, \chi_2, \chi_1', \chi_2'$, LHV theory gives the Bell-like bound [6,7]:

$$-1 \leq S \leq 0 \tag{34}$$

It is sufficient to realize one experimental situation which violates the bounds in (34).

For example, for $\chi_1 = 0°$, $\chi_2 = 67°$, $\chi_1' = 135°$, $\chi_2' = 23°$, (22), (27), (28), as obtained from QED, give $S = 0.207$ which violates (34) from above. For $\chi_1 = 0°$, $\chi_2 = 23°$, $\chi_1' = 45°$, $\chi_2' = 67°$, we obtain $S = -1.207$ violating (34) from below.

*Process 2:*

Here we put the two detectors on opposite sides of the $z$–axis. We consider all repeated experiments with pairs $e^+e^-$ produced in flight in the c.m. (located at the origin), each particle moving with speed $\upsilon = \beta c$, corresponding to all possible orientations of the axis along which a given pair moves, Here we must average over all angles $\theta, \phi$ of the vector $\vec{p}$, with $\vec{k}$ along the $z$-axis.

In the present case

$$\vec{k} = |\vec{k}|(0,0,1) \tag{35}$$

$$\vec{p} = |\vec{p}|(\cos\phi\sin\theta, \sin\phi\sin\theta, \cos\theta) \tag{36}$$

$$\vec{e}_1 = (-\cos\chi_1, \sin\chi_1, 0) \tag{37}$$

$$\vec{e}_2 = (-\cos\chi_2, \sin\chi_2, 0) \tag{38}$$

(see (7)) and

$$\vec{e}_1 \cdot \vec{p} = -|\vec{p}|\sin\theta\cos(\phi+\chi_1) \tag{39}$$

$$\vec{e}_2 \cdot \vec{p} = -|\vec{p}|\sin\theta\cos(\phi+\chi_2) \tag{40}$$

thus obtaining for (11)

$$P_r \propto \frac{[1 - 4(1-\beta^2)\cos(\phi+\chi_1)\cos(\phi+\chi_2)(\cos(\chi_1-\chi_2) - 2\cos(\phi+\chi_1)\cos(\phi+\chi_2))]}{(1-\beta^2\cos^2\theta)}$$

$$-\frac{4(1-\beta^2)^2\cos^2(\phi+\chi_1)\cos^2(\phi+\chi_2)}{(1-\beta^2\cos^2\theta)^2} - [\cos(\chi_1-\chi_2) - 2\cos(\phi+\chi_1)\cos(\phi+\chi_2)]^2$$

(41)

Upon using the integrals

$$\int_0^{2\pi} d\phi \cos(\phi+\chi_1)\cos(\phi+\chi_2) = \pi\cos(\chi_1-\chi_2) \tag{42}$$

$$\int_0^{2\pi} d\phi \cos^2(\phi+\chi_1)\cos^2(\phi+\chi_2) = \frac{\pi}{4}[1 + 2\cos^2(\chi_1-\chi_2)] \tag{43}$$

and

$$\int_0^\pi \frac{\sin\theta d\theta}{(1-\beta^2\cos^2\theta)} = \frac{1}{\beta}\ln\left(\frac{1+\beta}{1-\beta}\right) \tag{44}$$

$$\int_0^\pi \frac{\sin\theta d\theta}{(1-\beta^2\cos^2\theta)^2} = \frac{1}{\beta}\left[\frac{\beta}{1-\beta^2} + \frac{1}{2}\ln\left(\frac{1+\beta}{1-\beta}\right)\right] \tag{45}$$

with the latter two deduced from (14), (15), by replacing $\delta$ by $\pi/2$, we obtain

$$\int d\Omega P_r \propto A(\beta) + B(\beta)\cos^2(\chi_1-\chi_2) \tag{46}$$

where

$$A(\beta) = \frac{[4(2-\beta^2)-(1-\beta^2)^2]}{4\beta}\ln\left(\frac{1+\beta}{1-\beta}\right) - \frac{3}{2} + \frac{\beta^2}{2} \quad (47)$$

$$B(\beta) = -(1-\beta^2)\left[1 + \frac{(1-\beta^2)}{2\beta}\ln\left(\frac{1+\beta}{1-\beta}\right)\right] \quad (48)$$

and for the normalization factor we have upon summing over the set in (17),

$$N(\beta) = \frac{[4(2-\beta^2)-2(1-\beta^2)^2]}{\beta}\ln\left(\frac{1+\beta}{1-\beta}\right) - 8 + 4\beta^2 \quad (49)$$
$$\equiv 2[2A(\beta)+B(\beta)]$$

Accordingly, for the joint conditional probabilities, we have

$$P_\beta(\chi_1,\chi_2) = \frac{A(\beta)+B(\beta)\cos^2(\chi_1-\chi_2)}{2[2A(\beta)+B(\beta)]} \quad (50)$$

given that the two photons have emerged (back to back) along the $z$-axis.

For the measurement of only one of the polarizations, (50) leads to

$$P_\beta(\chi_1,\_) = \frac{A(\beta)+B(\beta)}{2[2A(\beta)+B(\beta)]}$$
$$= \frac{1}{2} = P_\beta(\_,\chi_2) \quad (51)$$

for all $0 \leq \beta \leq 1$, and the latter are, respectively, *independent* of $\chi_1, \chi_2$.

Again we have the important statistical property

$$P_\beta(\chi_1,\chi_2) \neq P_\beta(\chi_1,\_)P_\beta(\_,\chi_2) \qquad (52)$$

in general. It is interesting to note that an equality in (52) holds in the extreme relativistic case $\beta \to 1$ where each side is equal to $1/4$.

Only in the limiting case $\beta \to 0$, the joint probability in (50) for this process coincides with that in (22) for the first process.

As in (33), we define

$$S_\beta = P_\beta(\chi_1,\chi_2) - P_\beta(\chi_1,\chi_2') + P_\beta(\chi_1',\chi_2) + P_\beta(\chi_1',\chi_2') - P_\beta(\chi_1',\_) - P_\beta(\_,\chi_2) \qquad (53)$$

for four angles $\chi_1, \chi_2, \chi_1', \chi_2'$, and LHV theory gives [6,7]

$$-1 \leq S_\beta \leq 0 \qquad (54)$$

For $\beta \to 1$, an equality holds in (52), $S_\beta \to -1/2$, and this process, to be useful for testing the violation of (54), should not be conducted at very high speeds. For $\chi_1 = 0°$, $\chi_2 = 67°$, $\chi_1' = 135°$, $\chi_2' = 23°$, we have  for ,respectively, violating (54) from above, For $\chi_1 = 0°$, $\chi_2 = 23°$, $\chi_1' = 45°$, $\chi_2' = 67°$, we have $S_\beta = -1.12, -1.184, -1.201, -1.207$ for $\beta = 0.2, 0.1, 0.05, 0.01$ respectively, violating (54) from below. For $\beta$ larger than 0.2 but close to it, $S_\beta$ already turns out to be too close to the critical interval given in (54) to be relevant experimentally.

## 3. CONCLUTION

We have derived explicit closed expressions for joint probability distributions of photon polarizations correlations and for single photon polarization measurement of $\gamma\gamma$ in $e^+e^-$ annihilation in flight in two processes in QED. The mere fact that this quantum dynamical and ever reliable theory predicts a clear violation of the Bell-like inequality (34)/(35) for all $\beta$ for the first process and for several speeds, which are nevertheless high enough, for the second process, makes it interesting to carry out these experiments for the annihilation of free $e^+e^-$ in flight. Perhaps, such experiments may be easier to carry out than those involve with positronium decay [12] and we hope that this work will be of interest to both theoreticians and experimentalist alike.

ACKNOWLEDGEMENT

The authors would like to express the thanks for being granted a "Royal Golden Jubilee Award" by the TRF (Grant No. PHD/0022/2545) for the purpose of carrying out this project.